\documentclass[aps,prl,twocolumn,groupedaddress]{revtex4-1}

\usepackage{graphicx}
\usepackage{longtable}
\usepackage{booktabs} 
\usepackage{array} 
\usepackage{geometry}
\usepackage{marginnote}

\newlength{\onecolfig}
\newlength{\twocolfig}
\newcommand{\ion}[2]{\mbox{$^{#2}$#1$^+$}}
\newcommand{\Ca}[1]{\ion{Ca}{#1}}
\newcommand{\lev}[2]{\mbox{#1$_{\mbox{\tiny$#2$}}$}}

\newcommand{\hfslev}[3]{\mbox{#1$^{\mbox{\tiny$#3$}}_{\mbox{\tiny$#2$}}$}}

\newcommand{\unit}[1]{\,\mbox{#1}}

\newcommand{\Hz}{\unit{Hz}}
\newcommand{\kHz}{\unit{kHz}}
\newcommand{\MHz}{\unit{MHz}}
\newcommand{\GHz}{\unit{GHz}}

\newcommand{\mW}{\unit{mW}}

\newcommand{\mrad}{\unit{mrad}}

\newcommand{\um}{\unit{$\mu$m}}
\newcommand{\nm}{\unit{nm}}

\newcommand{\s}{\unit{s}}

\newcommand{\ms}{\unit{ms}}

\newcommand{\G}{\unit{G}}
\newcommand{\mG}{\unit{mG}}

\newcommand{\ish}{\mbox{$\sim$}\,}
\newcommand{\ltish}{\protect\raisebox{-0.4ex}{$\,\stackrel{<}{\scriptstyle\sim}\,$}}

\newcommand{\ket}[1]{\mbox{$\left| #1 \right>$}}
\newcommand{\sub}[1]{\mbox{$_{\mbox{\tiny #1}}$}}
\begin{document}

\title{High-fidelity spatial and polarization addressing of \Ca{43} qubits using near-field microwave control}

\author{D.~P.~L.~Aude~Craik}
\author{N.~M.~Linke}
\author{M.~A.~Sepiol}
\author{T.~P.~Harty}
\author{J.~F.~Goodwin}
\author{C.~J.~Ballance}
\author{D.~N.~Stacey}
\author{A.~M.~Steane}
\author{D.~M.~Lucas}
\author{D.~T.~C.~Allcock}\altaffiliation{Present address: National Institute of Standards and Technology, 325 Broadway, Boulder, CO 80305, USA}
\affiliation{Clarendon Laboratory, Department of Physics, University of Oxford, Parks Road, Oxford, OX1 3PU, UK}

\begin{abstract}
Individual addressing of qubits is essential for scalable quantum computation. Spatial addressing allows unlimited numbers of qubits to share the same frequency, whilst enabling arbitrary parallel operations. We demonstrate addressing of long-lived \Ca{43} ``atomic clock'' qubits held in separate zones ($960\um$ apart) of a microfabricated surface trap with integrated microwave electrodes. Such zones could form part of a ``quantum CCD'' architecture for a large-scale quantum information processor. By coherently cancelling the microwave field in one zone we measure a ratio of Rabi frequencies between addressed and non-addressed qubits of up to 1400, from which we calculate a spin-flip probability on the qubit transition of the non-addressed ion of $1.3\times 10^{-6}$. Off-resonant excitation then becomes the dominant error process, at around $5 \times 10^{-3}$. It can be prevented either by working at higher magnetic field, or by polarization control of the microwave field. We implement polarization control with error $2 \times 10^{-5}$, which would suffice to suppress off-resonant excitation to the $\sim 10^{-9}$ level if combined with spatial addressing. Such polarization control could also enable fast microwave operations.
\end{abstract}

\maketitle
\section{Introduction}
The prospect of a quantum computer (QC) that can solve classically intractable problems and make possible the simulation and engineering of quantum systems at the atomic level has driven research into many different candidate systems~\cite{Ladd2010}. The feasibility of large-scale quantum computation depends both on our ability to scale-up prototype systems to parallelized multi-qubit architectures and to perform qubit operations with error rates that are low enough to allow for the implementation of quantum error correction protocols without the need for a prohibitively large number of additional error-correcting qubits per logic qubit. Recent work using surface code error correction places the error threshold for fault-tolerant quantum computation as high as $\approx 1\%$, but error rates substantially below this are necessary for realistic overheads~\cite{Steane2003, Knill2005, Fowler2012}.

Almost all of the present QC technologies in which two-qubit quantum logic gates have been demonstrated are based on qubits which operate in the microwave frequency domain, for example laser-cooled trapped ions~\cite{Turchette1998} and neutral atoms~\cite{Saffman2015} based on hyperfine transitions, and cryogenically-cooled systems such as superconducting circuits~\cite{Barends2014, Devoret2013}, diamond or silicon defects~\cite{Awschalom2013,Childress2013}, and semiconductor quantum dots~\cite{Veldhorst2015}. In several of these systems, microwave-driven single-qubit rotations have been demonstrated with error rates well below the 1\% level~\cite{Brown2011,Shappert2013, Barends2014,Harty2014,Veldhorst2015}. Two-qubit quantum logic gates with error rates below 1\% have been achieved in superconducting circuits~\cite{Barends2014} and trapped ions ~\cite{Benhelm2008,Ballance2015b,Gaebler2016,Harty2016}. A critical issue for many of these technologies is the suppression of crosstalk between microwave signals targeted at specific qubits; even if the qubits operate at different frequencies, the limited bandwidth available means that frequency-sharing will ultimately be necessary. 

Trapped atomic ions are one of the most promising QC platforms~\cite{Monroe2013}, especially when hyperfine ground-level ``atomic clock'' qubits are used. These are free from spontaneous emission errors, offer very long coherence times (limited only by technical considerations)~\cite{Langer2005,Harty2014} with high-fidelity state preparation and readout~\cite{Myerson2008, Harty2014}, and can be directly manipulated by microwave radiation~\cite{Mintert2001, Ospelkaus2008, Ospelkaus2011, Khromova2012, Harty2016, Weidt2016} as well as by lasers~\cite{Leibfried2003, Kirchmair2009, Ballance2015b}. Prospects for scalability of trapped-ion microwave qubits were improved by the development of microfabricated ``chip'' ion traps with integrated microwave electrodes to provide all-electronic control of coherent qubit operations~\cite{Ospelkaus2011, Allcock2013,Warring2013PRA}. Two important challenges remain: performing microwave-driven two-qubit gates with sufficiently high fidelity, and demonstrating a scalable method of individually addressing ions using microwaves with sub-threshold crosstalk; we describe work towards the former goal in~\cite{Harty2016}, while the latter is the subject of this paper. These techniques could be combined with recent work on large multi-zone trap arrays \cite{Amini2010}, shuttling of ions between zones \cite{Blakestad2009, Moehring2011, Walther2012}, and sympathetic cooling using auxiliary ion species \cite{Barrett2003, Home2009}, to realize a ``quantum CCD'' architecture for a microwave-based quantum processor.

\begin{figure*} 
\centering
\includegraphics[width=0.8\textwidth]{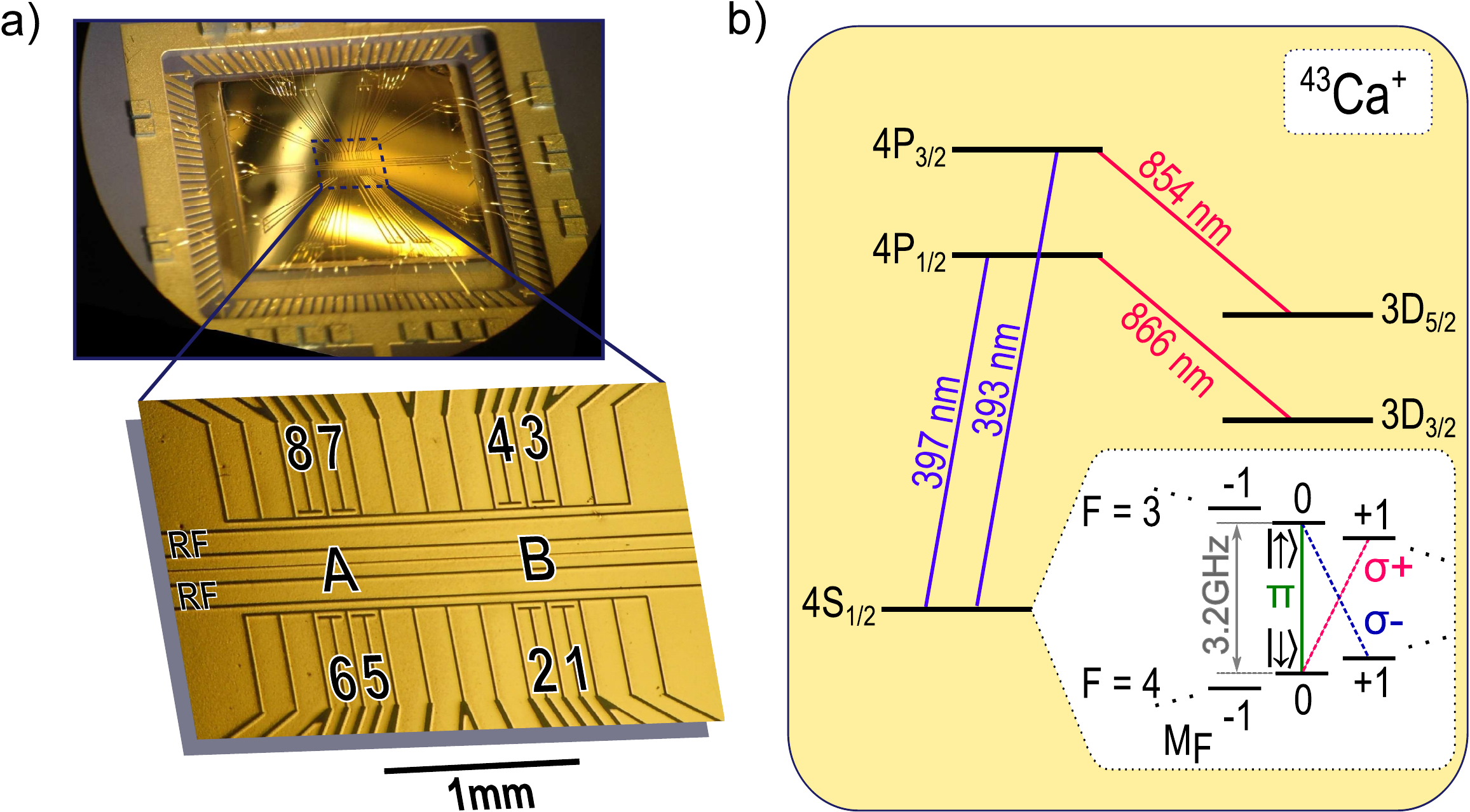}
\caption{\label{trapqubit}%
The surface ion trap and \Ca{43} transitions. 
(a) Top: ion trap chip mounted on a ceramic pin grid array (CPGA). Square gold-chip capacitors seen on the edge of the CPGA are used on DC electrodes (to short any microwave pick-up to ground) and on microwave electrodes (to allow DC voltages to be applied to the same electrodes). 
Bottom: magnified view of the centre of the device showing the separate trap zones A and B. The eight microwave electrodes (numbered) feature `T'-shaped slots around which microwave current flows. The axial electrodes used for the Paul trap radio-frequency (RF) confinement are indicated. A split centre DC electrode is used to control the orientation of the radial trap axes~\cite{Allcock2010}. 
(b) \Ca{43} energy levels, showing the four optical transitions used for laser cooling, and for qubit state preparation and readout. The inset shows relevant states within the ground level hyperfine structure, with the 3.2\GHz\ $\pi$ polarized qubit transition highlighted (green). $\sigma^+$ and $\sigma^-$ polarized transitions out of the qubit states used in the polarization-control experiment are also indicated (red and blue). Zeeman splittings between adjacent $M_F$ states are $\approx 0.98$\MHz\ at $B_0 = 2.8$\G.
}
\end{figure*}

Both far-field and near-field microwaves have previously been used to address trapped ions individually. In the far field, the long wavelength means that differential field amplitudes cannot be established between neighbour ions and addressing is instead done in frequency space, using a static magnetic field gradient to Zeeman-shift differentially the resonance frequencies of individual ions~\cite{Timoney2011, Piltz2014, Lake2014}. This scheme has achieved crosstalk errors of order $10^{-5}$ and is suited to a 1D linear array of magnetic-field-dependent qubits. 

Spatial addressing, which allows the use of noise-immune magnetic-field-independent ``atomic clock'' qubits, and 2D or 3D arrays of ions, is made possible if near-field microwaves are used since, in the near field of current-carrying waveguides, the field decay lengthscales depend on the geometry of the waveguides and of the surrounding structures, rather than on the wavelength of the radiation. It is feasible, with a single electrode of width $w$, to generate differential field amplitudes that can be used to address ions spaced by a few $w$ (provided induced return currents are carefully routed), while interfering fields from several such electrodes can produce field differentials at even smaller length scales.

Warring {\em et al.}~\cite{Warring2013PRL} demonstrated near-field addressing by selectively moving one of two ions in a single-zone trap on and off the null of a large-gradient quadrupole microwave field (also used to perform two-qubit microwave gates~\cite{Ospelkaus2011}), which was generated by feeding high-power microwave signals into the trap electrodes. That scheme achieved crosstalk errors of order $10^{-3}$; it would require the addition of further microwave nulling electrodes, similar to those used in this work, to enable parallel addressing in a 2D array of traps.

In this paper, we present experimental progress towards the implementation of the scheme we proposed in~\cite{AudeCraik2014} for microwave addressing of field-independent trapped-ion qubits.  The scheme is a microwave-driven implementation of the memory section of the quantum charge-coupled device (QCCD) architecture (proposed in \cite{Wineland1998} and further discussed in \cite{Kielpinski2002}) and involves storing single ions in separate trap zones where individually-addressed single-qubit rotations are applied in parallel using low-power microwaves. In this architecture, two-qubit gates are performed by shuttling two or more ions to entanglement zones where they are held in the same trap so that Coulomb-mediated gates can be performed via a shared motional mode. Although not implemented in this work, the ion shuttling required for this scheme has been performed in several surface-electrode traps with minimal motional heating \cite{Moehring2011, Fallek2016}. Here, we focus on implementing the large microwave field differentials needed to perform microwave-driven spatial addressing of qubits in neighbour zones.

\section{Spatial addressing experiment}
The demonstration experiments described here are performed in a prototype microfabricated surface-electrode dual-zone ion trap designed to provide arbitrary phase, amplitude and polarization control of the microwave field at each trap zone. Fabricated in Oxford cleanroom facilities using methods similar to those described in \cite{Allcock2010, AudeCraik2014}, the ion trap used in this work is a gold on fused-silica surface-electrode chip (fig.~\ref{trapqubit}a). The zone spacing of $960\um$ ($\approx 9$ ion-electrode distances) was chosen to facilitate the generation of large zone-to-zone microwave-field differentials, but making it smaller than a few ion-electrode distances would also hinder our ability to easily generate the electric potentials required for inter-zone splitting and shuttling, should we choose to implement them. The trap features two microwave addressing zones (denoted A and B), each with four integrated microwave control electrodes, which can be used to drive coherent single-qubit operations on an ion trapped $110\um$ from the chip surface. Crosstalk fields in neighbouring zones are nulled by generating cancelling fields using the microwave electrodes in those zones. The scheme could be scaled up to large arrays of trap zones, each of which could be addressed independently and in parallel. Further details relating to trap design, fabrication and the scalability of the addressing scheme can be found in \cite{AudeCraik2014, AudeCraikTh}. 

Using the Doppler recool method \cite{Wesenberg2007}, we measured an axial-mode heating rate for an ion trapped in between zones A and B of $\ish10^5$\,phonons/s, for an axial trap frequency of $0.5$\MHz.

We use the transition at frequency $f=3.226$\GHz\ between the $\ket{\downarrow}=\hfslev{4S}{1/2}{4,0}$ (where the superscript in the atomic term denotes the $F$, $M_F$ quantum numbers of the level) and $\ket{\uparrow}=\hfslev{4S}{1/2}{3,0}$ hyperfine ground-state levels of \Ca{43} as our qubit transition (fig.~\ref{trapqubit}b). At low static fields (here, $B_{0} = 2.8$\G), this transition is robust to magnetic field fluctuations (${\text{d}f}/{\text{d}B_{0}}=6.8\,$Hz/mG; ambient lab field fluctuations are of order 1\mG). For increased coherence times, intermediate-field clock qubits can also be used (for example, $\hfslev{4S}{1/2}{4,0}\leftrightarrow\hfslev{4S}{1/2}{3,+1}$ at $B_0=146\G$~\cite{Harty2014}, or $\hfslev{4S}{1/2}{4,+1}\leftrightarrow\hfslev{4S}{1/2}{3,+1}$ at $B_0=288$\G).  We prepare $\ket{\downarrow}$ using a $397\nm$ $\pi$-polarized beam with $3.2\GHz$ sidebands to optically pump on the $\hfslev{4S}{1/2}{3},\hfslev{4S}{1/2}{4}\leftrightarrow\hfslev{4P}{1/2}{4}$ transitions (the selection rule that forbids $M_F=0 \rightarrow M'_F=0$ when $\Delta F=0$ causes population to build up in the \ket{\downarrow} state). Readout is performed by first using a $\sigma^+$ circularly-polarized $393\nm$ beam to optically pump population from \ket{\downarrow} to the metastable $\hfslev{3D}{5/2}{}$ ``shelf'' state via \hfslev{4P}{3/2}{5}. The state is then read out using the $397\nm$ and $866\nm$ Doppler cooling lasers to look for fluorescence on the $\hfslev{4S}{1/2}{}\leftrightarrow\hfslev{4P}{1/2}{}$ $397\nm$ transition (the population transferred from \ket{\downarrow} to the shelf will not fluoresce, whilst any population remaining in \hfslev{4S}{1/2}{} will). This preparation and readout scheme, chosen here for its technical simplicity, achieves $\approx 90\%$ net state preparation and measurement (SPAM) fidelity. If desired, this can be straightforwardly increased to $\gtrsim 99.8\%$ using an extra laser at $850\nm$ (to repump population from $\hfslev{3D}{3/2}{}$ to $\hfslev{4P}{3/2}{}$) and microwave pulses, as we implemented in~\cite{Myerson2008,Harty2014}.

\begin{figure*}[ht!]
\centering
\includegraphics[width=0.84\twocolfig]{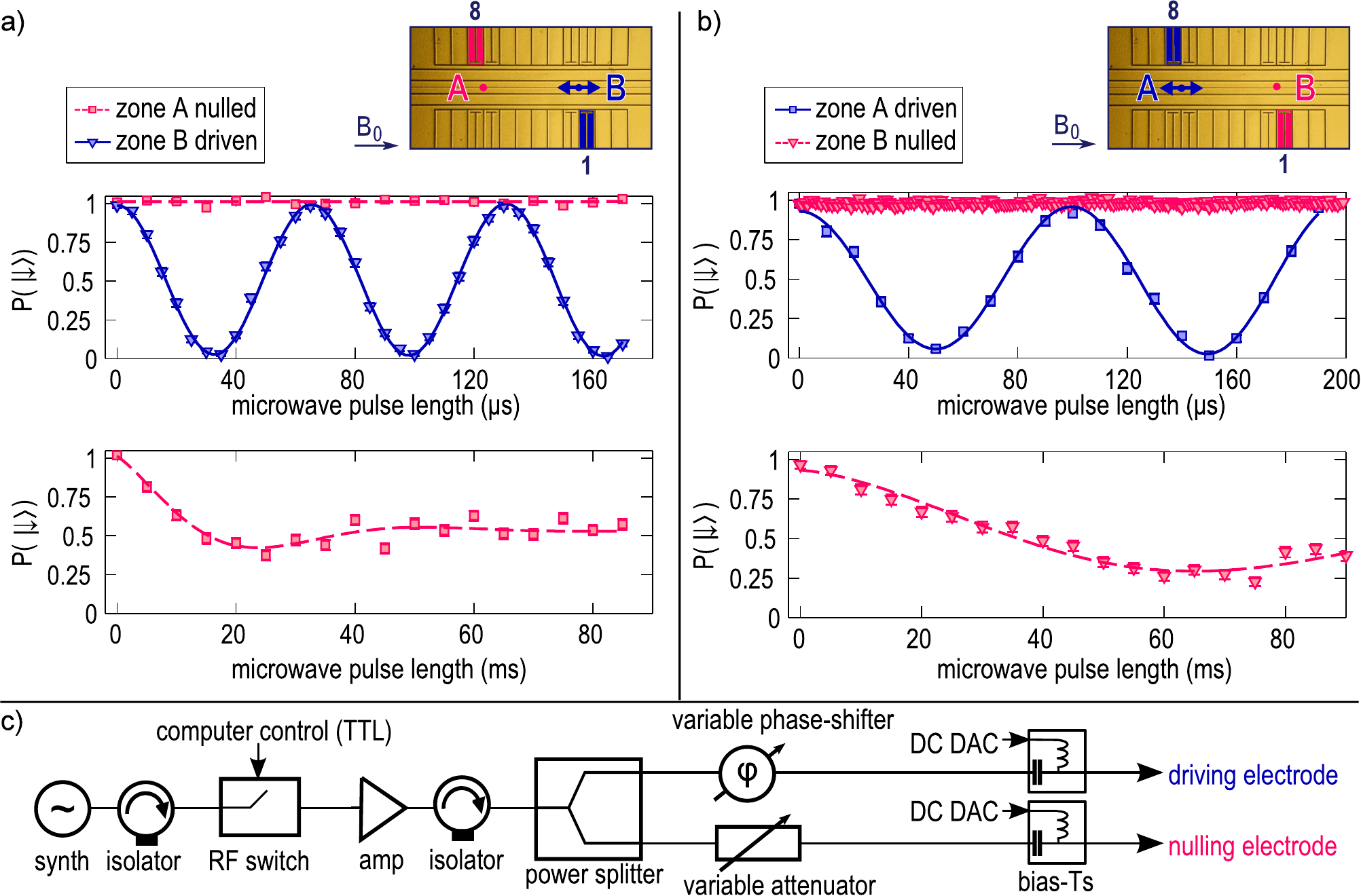}
\caption{\label{addressing}%
Spatial addressing experiments. %
(a) Zone B is the `driven' zone and zone A is the `nulled' zone. Top: Rabi flops in the driven zone (blue points, solid curve) with  $\Omega\sub{driven}=15.29(2)\kHz$ are seen when we scan the duration of a microwave pulse fed into electrodes 1 and 8. On this time-scale, the ion in the nulled zone is unaffected (red points, dashed line). Bottom: for longer pulse lengths, we are able to measure a small Rabi frequency in the nulled zone of  $\Omega\sub{nulled}=18(2)\Hz$ before decoherence reduces the contrast of the Rabi flops. %
(b) Similarly, with zone A as the driven zone and zone B as the nulled zone, we measure $\Omega\sub{driven}=10.07(3)\kHz$ and $\Omega\sub{nulled}=7.2(3)\Hz$. In both (a) and (b), the contrast of the Rabi flops has been normalized to correct for the $\approx$10\% state preparation and measurement errors. 
(c) The microwave system used to drive the two electrodes. The power levels at the vacuum feed-through for the data in plot (b) were $\approx170\mW$ at the driving electrode (8) and $\approx9\mW$ at the nulling electrode (1).}%
\end{figure*}

Single-qubit rotations are achieved by applying 3.2\GHz\ signals to the integrated microwave trap electrodes. Consider driving a single microwave electrode $i$, with all other electrodes 50\,ohm terminated. We define $\Omega\sub{near}$ to be the Rabi frequency measured on the qubit transition when the ion is trapped in the zone where the electrode is situated and $\Omega\sub{far}$ to be that measured when the ion is trapped instead in the neighbour zone. Thus $r_i=\frac{\Omega\sub{near}}{\Omega\sub{far}}$ is the Rabi frequency ratio produced between trap zones by a single microwave electrode $i$. Measured values for $r_i$ for all eight microwave electrodes are given in table~\ref{table:exptRabi}.

To demonstrate spatial addressing, we use a single electrode in the zone where we wish to address qubits (the ``driven zone'') to drive Rabi flops on an ion trapped in that zone. A second electrode located in the neighbouring ``nulled zone'' is used to null the microwave field along the spatial direction that couples to the qubit transition in that zone. We choose to use electrodes 1 and 8 because these produce the largest single-electrode Rabi frequency ratios ($r_1=3.6$ and $r_8=2.9$, see table~\ref{table:exptRabi}). The nulling is achieved by driving Rabi flops on the qubit transition, and minimizing the observed Rabi frequency as a function of the relative phase and amplitude of the microwave signals fed to the two electrodes (fig.~\ref{addressing}c). With nulling optimized, we then measure the qubit Rabi frequencies in both the driven and nulled zones, without further adjustment of the phase and amplitude. We achieve a Rabi frequency ratio between nulled and driven zones of $R_A=\frac{\Omega_{\text{nulled}}}{\Omega_{\text{driven}}}=1.2(1)\times 10^{-3}$ when zone A is the nulled zone (fig.~\ref{addressing}a), or $R_B=7.2(3)\times 10^{-4}$ when zone B is the nulled zone (fig.~\ref{addressing}b). Following \cite{Warring2013PRL}, we quantify the addressing error by the probability of exciting a spin-flip on the qubit transition of the neighbour ion when we drive a $\pi$-pulse (spin-flip) on the addressed ion. With this definition, these ratios imply an addressing error $\epsilon_A=\frac{\pi^2}{4} {R_A^2}=3.4(7)\times 10^{-6}$ and $\epsilon_B=1.27(7)\times 10^{-6}$ for zone A nulled and zone B nulled respectively.

\begin{figure}
\centering
\includegraphics[width=0.9\onecolfig]{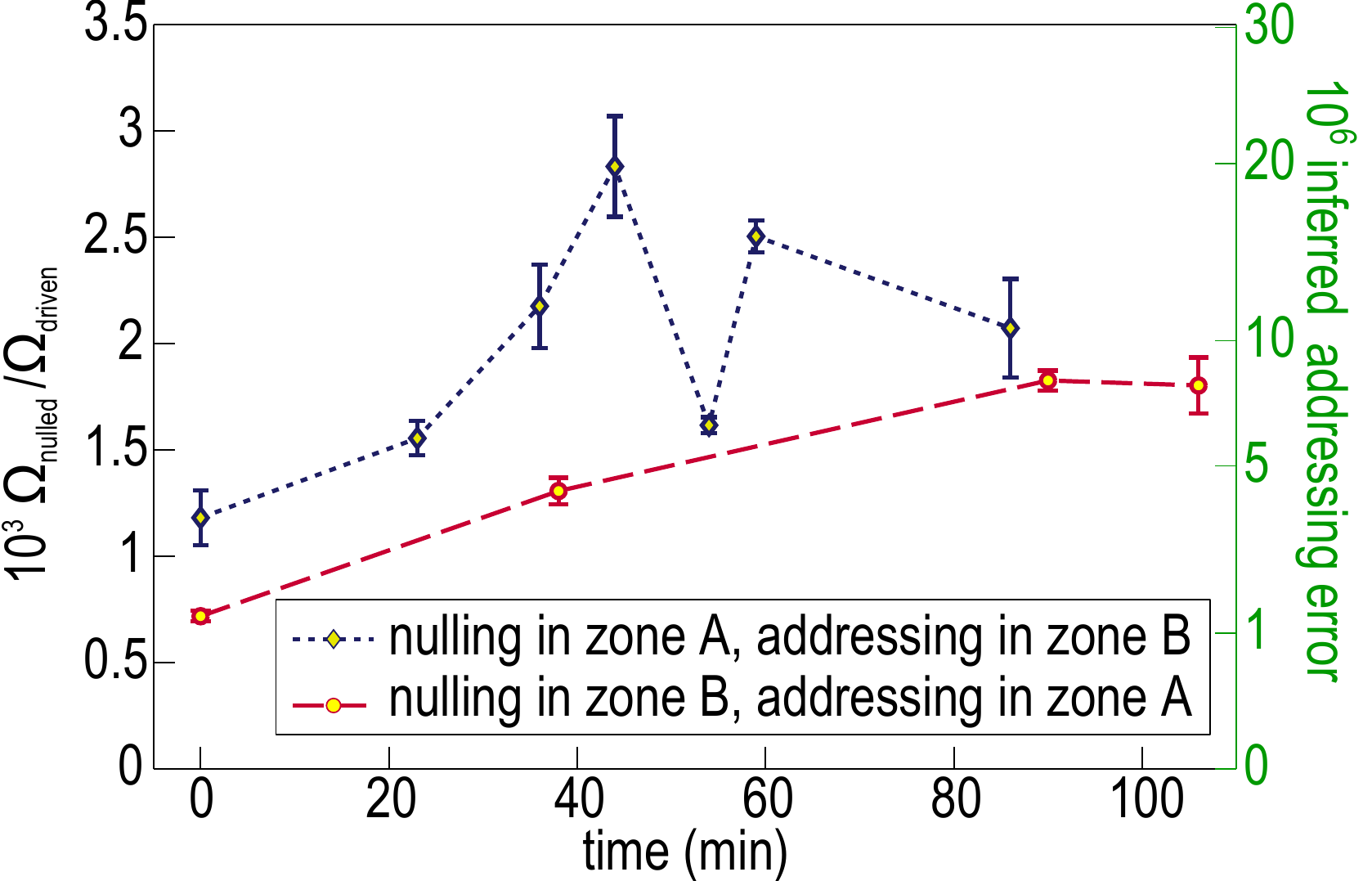}
\caption{\label{drift} The Rabi frequency in the `nulled' zone,  $\Omega_{\text{nulled}}$, was monitored for more than an hour. For each zone, the nulling parameters were optimized before $t=0$; thereafter no further adjustments to the nulling parameters were made. The ratio $({\Omega_{\text{nulled}}}/{\Omega_{\text{driven}}})$ is plotted, assuming constant $\Omega_{\text{driven}}$, for nulling in zone A (blue) and in zone B (red). Both ratios remain below $3\times 10^{-3}$, which implies the spin-flip errors can be kept below $2\times 10^{-5}$ without the need to recalibrate nulling over these time-scales. Lines are to guide the eye.}
\end{figure}

However, we note that two additional error sources arise in these experiments because we did not null the $\sigma$ components of the microwave polarization, which couple to spectator transitions out of the qubit states: (i) a non-zero probability of off-resonant excitation to the neighbouring $M_F=\pm 1$ states in the \lev{4S}{1/2} manifold (fig.~\ref{trapqubit}b); and (ii) an AC Zeeman shift on the qubit transition. We also measured the Rabi frequencies of $\sigma$ transitions and, from these, we can estimate the order of magnitude of both errors. For the conditions of the present experiments, the off-resonant excitation error (i) is more significant, causing excitation out of the qubit manifold with probability estimated to be between $0.6\times 10^{-3}$ and $5\times 10^{-3}$ (depending on whether the $\sigma$ field components from each electrode interfere destructively or constructively). The high-resolution data measured in zone B (fig.~\ref{addressing}b) set an upper limit of $\approx 3 \times 10^{-3}$ on this error for that case. The differential AC Zeeman shift (ii) is at most $340\Hz$ (calculated by assuming the worst-case phase relationship between $\sigma$ components) leading to a phase error $\phi=70\mrad$ on the nulled qubit when a $\pi$-pulse is applied on the driven qubit. Provided the AC Zeeman shift is stable, this can be corrected for by keeping track of the qubit phase. Assuming similar stability of the $\sigma$ fields as found for the $\pi$ fields (fig.~\ref{drift}), we estimate fluctuations in the phase error $\delta\phi<1\mrad$, leading to a reduction of the qubit state fidelity $\ish (\delta\phi)^2 \ltish 10^{-6}$ which is negligible compared to the off-resonant excitation error. Both error sources can be straightforwardly reduced either by decreasing the Rabi frequency, shaping the microwave pulses in time, or by increasing the static magnetic field. In practice one would ideally wish to use an intermediate-field clock qubit such as that at $B_0=288\G$ where, without reduction in Rabi frequency, the off-resonant excitation error would be $\ish 5 \times 10^{-7}$ and the AC Zeeman shift $\ltish 4\Hz$ (see table~\ref{table:errorBudget}). Alternatively, the unwanted polarization components of the field can also be nulled, as demonstrated below, which would reduce these errors to negligible levels even at low static fields and/or high Rabi frequencies. 

The decoherence seen in the ``nulled zone'' Rabi flops (fig.~\ref{addressing}) most likely occurs due to relative phase and amplitude fluctuations between the two arms of the microwave drive interferometer (which cause the minimized Rabi frequency to change from shot to shot of the experiment). We confirm that this decoherence is not a property of the qubit by performing a control experiment, where we eliminate relative phase and amplitude fluctuations by using a single microwave line to drive only one of the microwave electrodes with an attenuated signal so as to produce very low Rabi frequency (18\Hz). We observe no significant loss of contrast in a microwave pulse time scan up to 55\ms, which is consistent with the expected coherence time for these qubit states, elsewhere measured to be $T_2^*=6(1)$\s\ at $B_0=2.0\G$~\cite{Ballance2015b}. To investigate slow drifts in the nulling quality, we monitored the stability of the nulling over $>1$~hour in each zone (fig.~\ref{drift}) and found that both $R_A$ and $R_B$ remain below $3\times 10^{-3}$ over this period, sufficient to maintain $\epsilon_{A,B}\ltish 2\times 10^{-5}$ without recalibration of the nulling field.

\section{Polarization addressing experiment}
\begin{figure}[]
\centering
\includegraphics[width=0.9\onecolfig]{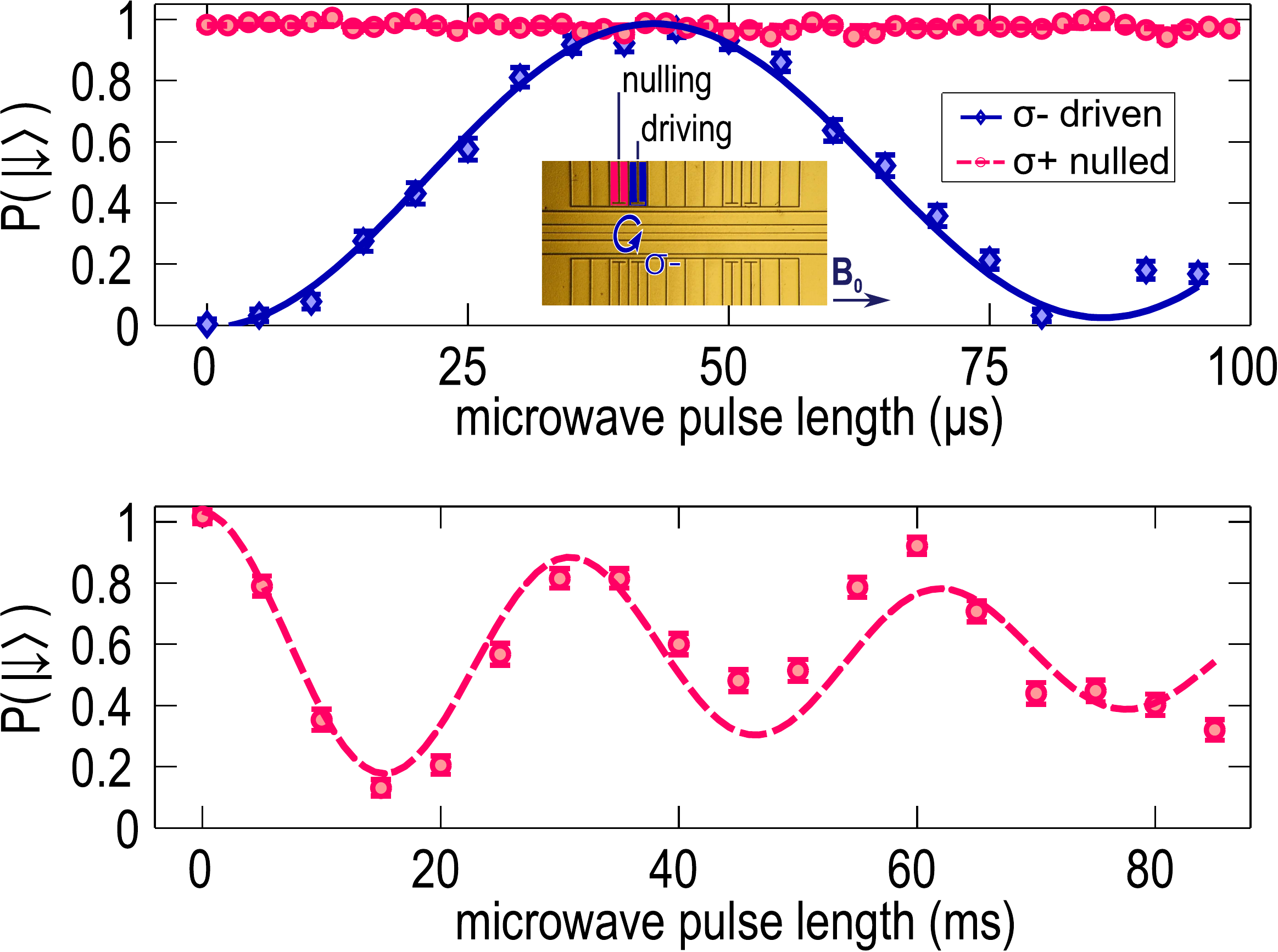}
\caption{\label{polarization} To demonstrate polarization control, we null the $\sigma^+$ component of the microwave polarization. We are then able to drive the $\sigma^-$-polarized transition that is indicated in blue on fig.~\ref{trapqubit}b, without exciting the nearly-degenerate $\sigma^+$-polarized transition indicated in red on  fig.~\ref{trapqubit}b. Top: we drive Rabi flops on the driven $\sigma^-$-polarized transition (blue points, solid curve) with Rabi frequency $\Omega_{\sigma^-}= 11.6(1)\kHz$ by scanning the length of the microwave pulse applied to electrodes 7 and 8. On this time-scale, the nulled $\sigma^+$-polarized transition is not visibly excited (red points). Both datasets on this plot were taken with microwaves resonant with the $\sigma^+$ transition (which is $1.6\kHz$ detuned from the  $\sigma^-$ transition).  Bottom: on the millisecond time-scale, we observe slow Rabi flops on the nulled $\sigma^+$-polarized transition with Rabi frequency $\Omega_{\sigma^+}= 32.0(8)\Hz$, driven by residual $\sigma^+$-polarization. Data have been corrected for the $\approx$10\% state preparation and measurement errors.}
\end{figure}

With four electrodes per zone, the ion trap used here gives us more than enough degrees of freedom to control the microwave field in all three spatial dimensions and hence to null unwanted polarization components. This could enable significantly faster microwave operations by eliminating the off-resonant excitation errors discussed above. It is also useful, for example, for performing high-fidelity read-out at low $B_0$, which involves selectively transferring population in one of the qubit states to the stretch state \hfslev{4S}{1/2}{4,+4}: the $\sigma^+$ and $\sigma^-$ transitions out of the qubit states are nearly degenerate in frequency (1.6\kHz\ splitting at $B_0=2.8\G$), hence off-resonant excitation limits the speed at which this transfer can be performed. A similar difficulty arises for readout of the $\hfslev{4S}{1/2}{4,+1}\leftrightarrow\hfslev{4S}{1/2}{3,+1}$ field-independent qubit at $B_0=288\G$. Here we demonstrate selective nulling of the $\sigma^+$ polarization component of the microwave field in zone A by using two microwave control electrodes in that zone (electrodes 7 and 8) and the same drive system used in the experiment described above. Nulling is achieved by adjusting the phase shifter and variable attenuator while minimizing the ion's Rabi frequency on the $\hfslev{4S}{1/2}{4,0}\leftrightarrow\hfslev{4S}{1/2}{3,+1}$ $\sigma^+$ transition. To measure the Rabi frequency on the $\hfslev{4S}{1/2}{4,+1}\leftrightarrow\hfslev{4S}{1/2}{3,0}$ $\sigma^-$ transition, we first prepare the $\ket{\uparrow}$ qubit state by optically pumping to $\ket{\downarrow}$ and then driving a microwave $\pi-$pulse on the qubit transition using a third electrode. We achieve a Rabi frequency ratio of ${\Omega_{\sigma^+}}/{\Omega_{\sigma^-}}=2.77(8)\times 10^{-3}$ (fig.~\ref{polarization}), implying a polarization addressing error of $\epsilon\sub{pol}=1.9(1) \times 10^{-5}$. Combining the polarization control with the spatial addressing would, for the typical 10\kHz\ Rabi frequencies used here, suppress the off-resonant transition errors out of the qubit states to negligible levels ($\ish 10^{-9}$), as the off-resonant $\sigma$ transitions are $\approx$1\MHz\ detuned from the qubit transition at $B_0=2.8\G$. 

\section{Future work and conclusions}
In considering what improvements can be made for a next-generation trap, we note the data in table~\ref{table:exptRabi} shows that five of this trap's microwave electrodes produced Rabi frequency ratios $r_i>1$ between addressing zones, but the remaining three did not. Large $r_i$ are desirable so that crosstalk between zones is local, improving the ease of scalability to a large array of zones. Numerical simulations predicted $r_i\ish 5$ for all microwave electrodes. We believe that the discrepancy between simulated and measured ratios may be due to the fact that the ions are not directly above the microwave electrodes, and hence are removed from the points where each electrode produces the greatest field strength. At the ion positions, the electrode fields have already decayed to approximately one-tenth of their maximum value and are then of the same order of magnitude as the fields produced by return currents coupled across the trap. The ratio of field amplitudes between the two zones is very sensitive to the precise distribution of these stray currents, which may be strongly affected by fabrication inaccuracies and any other differences between the simulated and fabricated trap (such as the CPGA package, or the electrical loading of the RF drive system, neither of which we attempted to simulate). 

We identify three critical improvements that should be incorporated into future near-field microwave addressing experiments that will allow for control of larger numbers of ions with lower levels of crosstalk. Firstly, the addressing electrodes need to be placed as close to the ion on the trap as possible (for example, within the centre control electrode). To do this for more than a few electrodes will require moving from a single-layer to a multi-layer electrode fabrication technology, for example as proposed in~\cite{AudeCraik2014, AudeCraikTh}. Secondly, a low crosstalk and impedance-controlled package for the trap should replace the CPGA. Thirdly, a microwave drive system with many digitally controlled channels is required so that calibration of field nulling can be automated.

\setlength{\extrarowheight}{2pt}
\begin{table}[!t]
\centering
\setlength{\tabcolsep}{3pt}
\begin{tabular}{|c|cccc|cccc|}
\hline
~~Electrode $i$~~ & 1 & 2 & 3 & 4 & 5 & 6 & 7 & 8 \\ \hline 
$r_i=\frac{\Omega\sub{near}}{\Omega\sub{far}}$ 
& $3.6$ & $0.93$ & $2.6$ & $1.7$ & $0.43$ & $1.5$ & $0.78$ & $2.9$ \\[2pt] \hline
\end{tabular}
\caption{\label{table:exptRabi}
Measured Rabi frequency ratios for all microwave drive electrodes. For each electrode $i$, the quantity $r_{i}$ gives the ratio of the Rabi frequency $\Omega\sub{near}$ measured for an ion trapped in the zone near electrode $i$, to the Rabi frequency $\Omega\sub{far}$ measured for an ion trapped in the zone far from electrode $i$. The Rabi frequencies are measured on the $\ket{\downarrow}\leftrightarrow\ket{\uparrow}$ qubit transition, for a static magnetic field applied parallel to the trap axis (as shown in figure~\ref{addressing}). Estimated measurement errors are $\approx8\%$. (If we instead align the static field perpendicular to the trap axis, we measure $r_i\approx 1$ for all electrodes, possibly because stray currents travelling down the RF electrode produce a $\pi$-polarized microwave field that drives the qubit transition with similar strength in both trap zones.)}
\end{table}

\begin{table*}[!t]
\centering
\caption{\label{table:errorBudget} Error budget table summarizing the metrics used in this work to assess the performance of the field-nulling addressing scheme. The second column details which metrics were directly measured and which were calculated from measurement. The last column of the table lists the projected errors for the scheme when implemented on an intermediate-field \Ca{43} clock qubit at 288\G\ (both the low-field clock transition used here and the 288\G\ transition are $\pi$-polarized). The 288\G\ qubit was not used in this work because it requires a magnetic field strength beyond the capabilities of the present apparatus. The $\sigma$-polarized 146\G\ clock qubit (previously used to implement high-fidelity single-qubit and two-qubit microwave-driven  gates \cite{Harty2014, Harty2016}) can also be used with this addressing scheme, with a projected total addressing error of order $10^{-5}$. All errors given here relate to addressing performed with a single microwave electrode per zone. As was separately demonstrated in this work, more electrodes per zone can be used to control the polarization of the microwave field, a technique which would reduce or eliminate off-resonant excitation and AC Zeeman shift errors if combined with addressing. The errors are quoted below for the nulled qubit, but similar levels of off-resonant excitation and light-shift also occur in the driven qubit.}
\begin{tabular}{m{4.8cm} m{0.1cm} m{4cm} m{0.1cm} m{2.2cm} m{0.1cm} m{2.9cm}}
\hline
Metric  && How was it evaluated? && Value for this experiment && Projected error for intermediate-field qubit at $B_0=288\G$\\
\hline
\hline
Ratio of qubit-transition Rabi frequencies, $R_{A,B}=\frac{\Omega\sub{nulled}}{\Omega\sub{driven}}$&& Measured&&$\leq 1.2(1)\times 10^{-3}$&& $\leq 1.2(1)\times 10^{-3}$ (same as for low-field qubit)\\
\hline 
Probability of spin-flip on the qubit transition of nulled qubit when we flip the spin of the driven qubit, $\epsilon_{A,B} = \frac{\pi^2}{4} {R_{A,B}^2}$&&Calculated from measured $R_{A,B}$&&$\leq 3.4(7)\times 10^{-6}$&&$\leq 3.4(7)\times 10^{-6}$ (same as for low-field qubit)\\

\hline
Off-resonant excitation error on nulled qubit&&Calculated from measured ratio of $\sigma$ to $\pi$ Rabi frequencies for electrodes 1 and 8 in each zone&&$\leq 5\times 10^{-3}$&&$\leq 5\times10^{-7}$\\
\hline
AC Zeeman shift on nulled qubit&&Calculated from measured ratio of $\sigma$ to $\pi$ Rabi frequencies for electrodes 1 and 8 in each zone&&$<340$\Hz&& $<4$\Hz \\
\hline
Phase error on the nulled qubit when we drive spin-flip on addressed ion&&Calculated from measured ratio of $\sigma$ to $\pi$ Rabi frequencies for electrodes 1 and 8 in each zone&&$<70$mrad&&$<0.7$mrad\\ 
\hline
Stability of phase error on nulled qubit && Estimated from measured ratio of $\sigma$ to $\pi$ Rabi frequencies for electrodes 1 and 8 in each zone and from measured drift of nulled qubit Rabi frequency&&$<1$mrad&&$<0.01$mrad\\ 
\hline
Total addressing error&&Measured from high-resolution data taken in zone B (see fig.~\ref{addressing}b top)&& $<3\times 10^{-3}$&&$\leq 4\times 10^{-6}$\\ 
\hline
\end{tabular}
\end{table*}
In conclusion, this work reports progress towards the implementation of a near-field microwave addressing scheme which can enable arbitrary operations to be performed in parallel on ``atomic clock'' memory qubits stored in different zones of a surface-electrode QCCD-type processor. The implementation of this scheme with high fidelity hinges on the achievement, reported here, of a large field differential between processor zones (which are spaced by $960\um$, a mere 1\% of the wavelength of the qubit-driving microwaves) using relatively low microwave powers and an electrode geometry which offers independent field control at each zone. Such independent control is achieved by interfering fields generated by microwave electrodes integrated into each trap zone so as to null residual crosstalk in neighbour zones. For technical simplicity, the nulling was tested here using a low-field \Ca{43} qubit at 2.8\G, for which the addressing error was limited by off-resonant excitation of spectator transitions at the $10^{-3}$ level. If an intermediate-field qubit (such as the 288\G\ clock qubit in \Ca{43}) is used instead, off-resonant excitation will be suppressed to below the $10^{-6}$ level and the field ratios of up to 1400 demonstrated here will produce a total addressing error of $<4 \times 10^{-6}$ (see table~\ref{table:errorBudget}). We also demonstrate polarization control of the microwave field with sufficient precision to suppress off-resonant transitions out of low-field qubit states to negligible levels; if combined with crosstalk-nulling, this polarization control would permit high-fidelity addressing to be performed even in low-field qubits. Furthermore, we verify that the field nulling used to suppress crosstalk between trap zones is passively robust to laboratory environmental fluctuations for over an hour, indicating that, after an initial calibration, it is feasible to perform single-qubit addressing operations with errors $\ltish 10^{-5}$ for long periods using this scheme. 

These error rates are two to three orders of magnitude below the threshold rates necessary for fault-tolerant quantum computation. The techniques used to null the qubit-qubit crosstalk here may be applicable to the wide variety of physical systems which use qubits based on microwave frequency transitions, and are likely to be essential if these systems are to be scaled up to the large numbers of qubits required for general-purpose quantum computing.

\section{Acknowledgments}
We thank D.~MacDougal for helpful {\tt MATLAB} scripts, L.~Stephenson for technical assistance, and C.~J.~Stevens, A.~D.~Karenowska and P.~J.~Leek for useful discussions. This work is supported by the U.S.\ Army Research Office (contract W911NF-14-1-0217), and by the U.K.\ EPSRC ``Networked Quantum Information Technology'' Hub (Grant No. EP/M013243/1).

\end{document}